\documentclass{nature}
\usepackage{latexsym}
\usepackage{amsthm}
\usepackage{amssymb}
\usepackage{amsmath}
\usepackage[all]{xy}
\usepackage{graphicx}
\usepackage{subfigure}
\usepackage{dcolumn}
\usepackage{bm}
\usepackage{bbm}
\usepackage[usenames,dvipsnames]{xcolor}
\usepackage{amsfonts}
\usepackage{cases}
\usepackage[a4paper,pagebackref=true,colorlinks=true,
linkcolor=blue,citecolor=blue,
pdfauthor={ },
pdftitle={ },
pdfsubject={ },
pdfkeywords={ }]{hyperref}

\begin{document}

\title{Sensing and atomic-scale structure analysis of single nuclear spin clusters in diamond}

\author{Fazhan Shi$^{1}$, Xi Kong$^{1}$, Pengfei Wang$^{1}$, Fei Kong$^{1}$, Nan Zhao$^{2}$, Ren-Bao Liu$^{3}$, Jiangfeng Du$^{1}$\thanks{e-mail: djf@ustc.edu.cn}}

\maketitle

\begin{affiliations}
\item
National Laboratory for Physics Sciences at the Microscale and Department of Modern Physics, University of Science and Technology of China, Hefei, 230026, China
\item
Beijing Computational Science Research Center, Beijing 100084, China
\item
Department of Physics and Centre for Quantum Coherence, The Chinese University of Hong Kong, Shatin, New Territories, Hong Kong, China

\end{affiliations}

\begin{abstract}
Single-molecule nuclear magnetic resonance (NMR) is a crown-jewel challenge in the field of magnetic resonance spectroscopy and has important applications in chemical analysis and in quantum computing. Recently, it becomes possible to tackle this grand challenge thanks to experimental advances in preserving quantum coherence of nitrogen-vacancy (NV) center spins in diamond as a sensitive probe\cite{Wrachtrup2008Nature, Lukin2008Nature, Wrachtrup2013Science, Rugar2013Science} and theoretical proposals on atomic-scale magnetometry via dynamical decoupling control\cite{Liu2011NatNano}. Through decoherence measurement of NV centers under dynamical decoupling control, sensing of single $^{13}\textbf{C}$ at nanometer distance has been realized\cite{Lukin2012PRL, Hanson2012PRL, Wrachtrup2012NatNano}. Toward the ultimate goal of structure analysis of single molecules, it is highly desirable to directly measure the interactions within single nuclear spin clusters. Here we sensed a single $^{13}\textbf{C}$-$^{13}\textbf{C}$ nuclear spin dimer located about 1 nm from the NV center and characterized the interaction between the two nuclear spins, by measuring NV center spin decoherence under various dynamical decoupling control. From the measured interaction we derived the spatial configuration of the dimer with atomic-scale resolution. These results demonstrate that central spin decoherence under dynamical decoupling control is a feasible probe for NMR structure analysis of single molecules.
\end{abstract}
\maketitle

In NMR, the structures of complex molecules can be unravelled by analysis of interaction between nuclear spins. Traditional magnetic resonance techniques rely on large ensembles of nuclear spins to accumulate signal-to-noise ratios. Recently, novel methods have been developed to improve the sensitivity including magnetic resonance force microscopy\cite{Rugar2009PNAS} and optically detected magnetic resonance (ODMR)\cite{Wrachtrup2008Nature, Lukin2008Nature, Yacoby2012NatNano, Degen2013PRL, Wrachtrup2013Science, Rugar2013Science}.
Non-interacting single nuclear spins have been successfully sensed using NV center probes under dynamical decoupling control.
However, measurement of interaction between nuclear spins within a single molecule and resolving their structures are still challenging.
Here we report experimental characterization of coherent coupling within a two-nuclear-spin cluster close to an NV center in diamond and determination of the cluster's spatial configuration using the interaction information.

The NV center consists of a nitrogen impurity and a neighbour vacancy.
Its triplet (S=1) ground state can be spin-polarized and read out optically.
It has ultralong coherence time up to milliseconds in ultrapure samples\cite{Wrachtrup 2009 NatMat}
or under dynamical decoupling control\cite{Cory2010PRL}.
These remarkable properties make the NV center a potential candidate for observing novel quantum phenomena\cite{Du2011NatComm, Liu2011PRL}, quantum processing\cite{Wrachtrup2006NatPhys, Lukin2007Science, Wrachtrup2008Science, Awschalom2009Science, Lukin2009PRL, Lukin2009Science, Wrachtrup2010NatPhys, Du2010PRL, Hanson2012Nature, Du2012PRL, Wrachtrup2013NatPhys}
and magnetic field sensing\cite{Lukin2008Nature, Wrachtrup2008Nature}.

In pure diamond samples, NV center electron spins lose their coherence due to the magnetic fluctuation from surrounding $^{13}$C nuclear spins.
The fluctuation is dominated by single nuclear spin precession or flip-flop processes of spins clusters.
Our measurement was carried out in strong field, where single spin precession was strongly suppressed\cite{Liu2012PRB},
and therefore the interacting $^{13}$C spin clusters are the dominant noise source.
In this case, we are able to single out the physical effect caused by collective dynamics of nuclear spin clusters and, then, unravel the interaction within the cluster.

Among different kinds of $^{13}$C nuclear spin clusters, a bonded $^{13}$C spin dimer stands out because it can induce strong fingerprint modulations on NV center spin coherence\cite{Liu2011NatNano}.
In this work, we observed the dimer-induced modulation on NV center electron spin coherence,
and characterized the interaction within the dimer and that between the dimer and the NV center electron spin.

In strong magnetic field, the dimer dynamics can be described by a pseudo-spin model,
where the two-spin states $|\uparrow\downarrow\rangle$ and $|\downarrow\uparrow\rangle$ of the dimer are mapped to the spin-up $\vert \Uparrow\rangle$ and spin-down $\vert \Downarrow\rangle$ states of the pseudo-spin $\boldsymbol{\sigma}$.
The polarized dimer states $|\uparrow\uparrow\rangle$ and $|\downarrow\downarrow\rangle$ are energetically separated and irrelevant in the dynamics because of the large Zeeman energy in the strong field.
The dynamics of the pseudo-spin is governed by the Hamiltonian\cite{Liu2011NatNano}
\begin{equation}
\label{pair Hamiltonian}
H_{ps}^{(m_S)}=\frac{1}{2}\textbf{h}^{(m_S)}\cdot
\mbox{\boldmath$\sigma$}=\frac{1}{2}( X\sigma_x+Z^{(m_S)}\sigma_z),
\end{equation}
where the component $X$ of the pseudo-field $\mathbf{h}^{(\alpha)}$ is the nuclear dipolar interaction strength
and the component $Z^{(m_S)}$ is due to the difference between hyperfine couplings of the two nuclei to the center spin, corresponding to the NV center electron spin state $|m_S\rangle$.

The pseudo-field component $Z^{(m_S)}$ and hence the nuclear spin pair dynamics depend on the NV center electron spin state $\vert m_S\rangle$.
For the NV center spin in $\vert m_S=0\rangle$ state, the hyperfine couplings vanish while for $\vert m_S=1\rangle$ state, the hyperfine couplings of the two nuclear spins have a finite difference of $Z^{(1)}\sim 10$~kHz.
Correspondingly the pseudo-spin will precess with frequencies $\omega_{\text{ps}}^{(0)}=X$ or $\omega_{\text{ps}}^{(1)}=\sqrt{X^2+\left[Z^{(1)}\right]^2}$.
When the NV center spin is prepared in a superposition state $\left(\vert m_S=0\rangle+\vert m_S=1\rangle\right)/\sqrt{2}$, the characteristic frequency of the noise induced by the pseudo-spin becomes $\omega_{\text{ps}}^{\text{eff}}=\sqrt{X^2+\left[\frac{Z^{(1)}}{2}\right]^2}$ (SI, Sec.~1.2).
In our experiment, the pseudo-spin precession frequency $\omega_{\text{ps}}^{\text{eff}}$ was measured by dynamical decoupling control\cite{Du2009Nature, Hanson2010Science, Cory2010PRL} of the NV center spin.

\begin{figure}
\centering
\includegraphics[width=0.95\columnwidth]{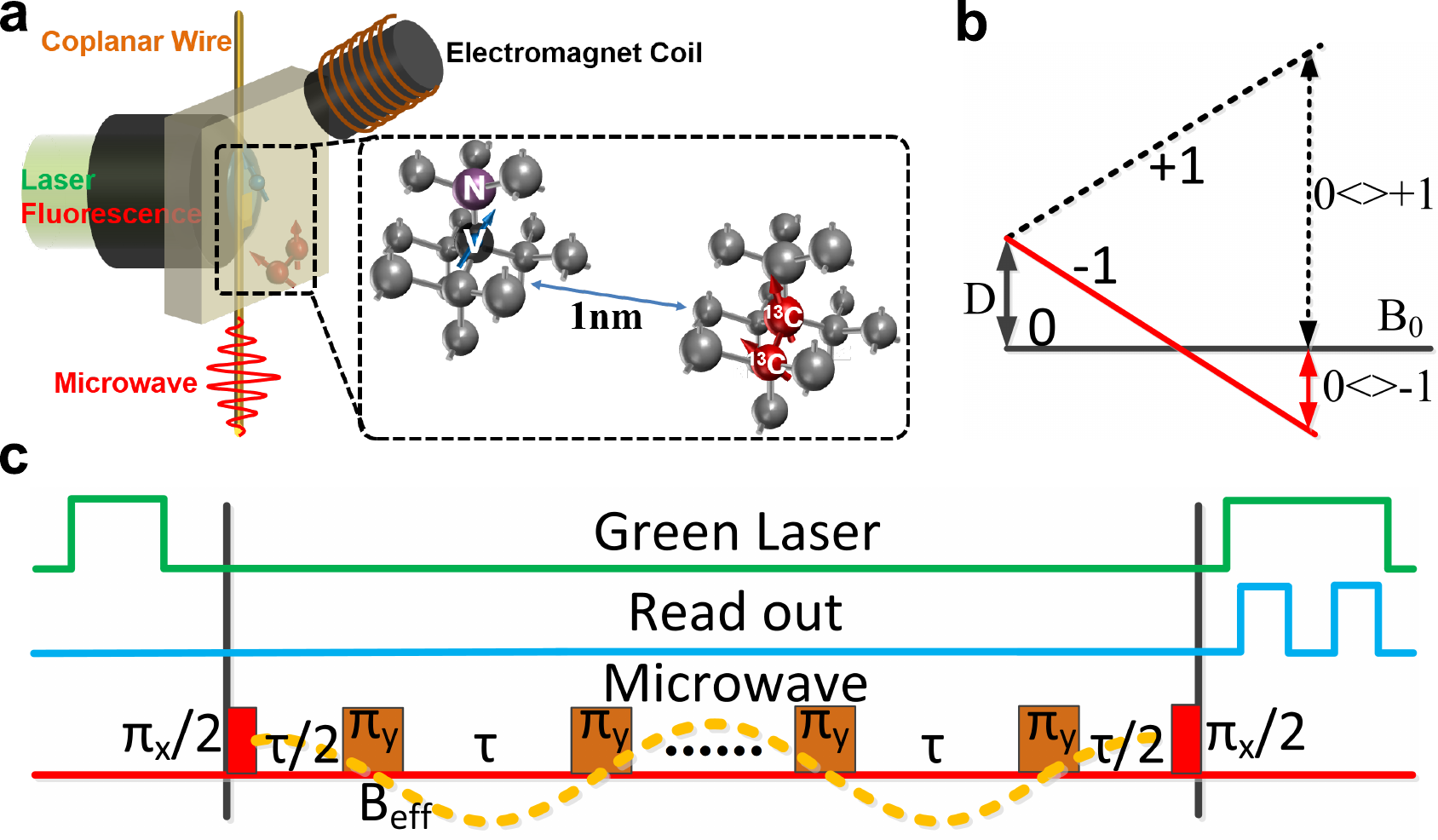}
    \caption{\textbf{Schematics of the setup and experimental methods.}
   (a) We measured the spin coherence of an NV sensor coupled to a $^{13}\text{C-}^{13}\text{C}$ dimer. The external DC magnetic field was applied by a movable electric magnet while microwave was carried by a coplanar wire.
   (b) The degeneracy of the NV center spin states $|\pm1\rangle$ was lifted by an external magnetic field B${_0}$ along the NV axis. We encoded the quantum transition 0$\leftrightarrow$-1 (or 0$\leftrightarrow$1) as a qubit and manipulated it by resonant microwave pulses.
   (c) The pulse sequences contained two green laser pulses for initialization and readout of the electron spin state and CPMG microwave pulses in between to control the spin. 
 }
    \label{setup}
\end{figure}

\begin{figure}
\centering
\includegraphics[width=0.95\columnwidth]{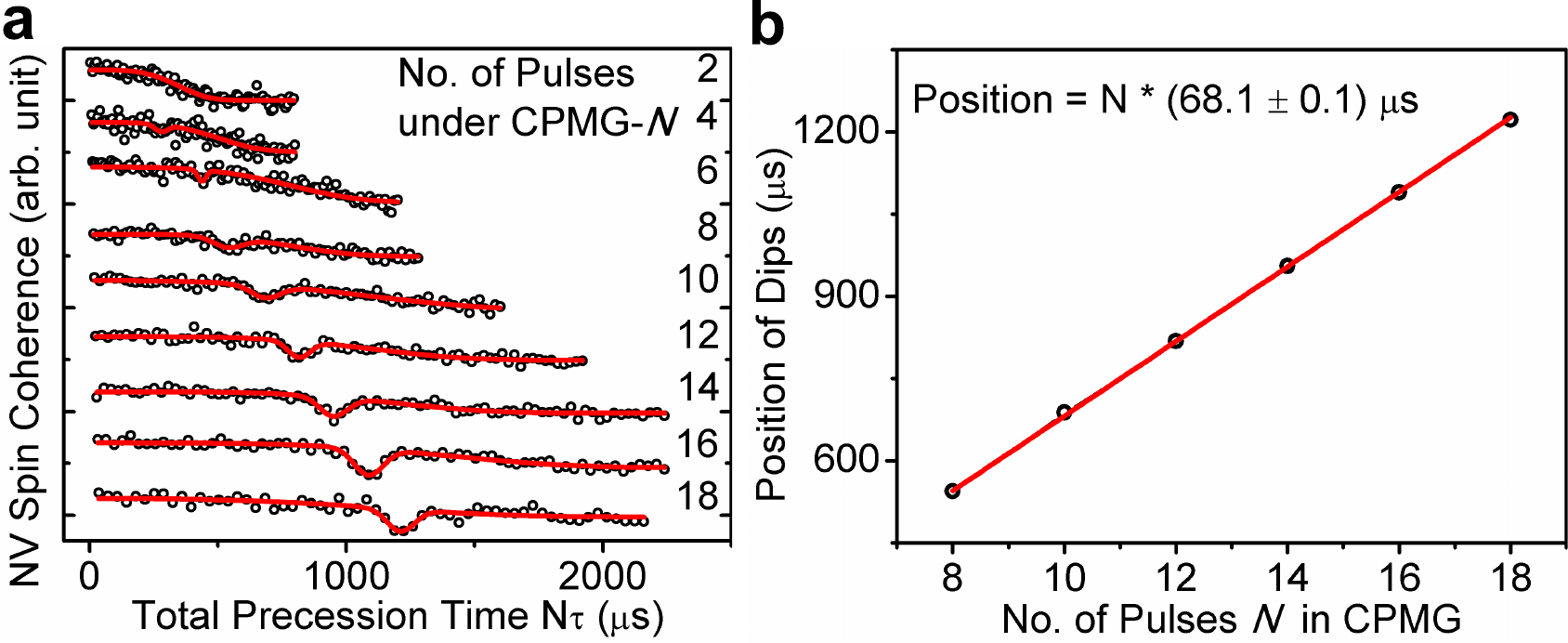}
    \caption{\textbf{Dip features in the NV center spin decoherence under CPMG control with various numbers of pulses.}
   (a) When the pulse number $N$ of CPMG increased from 2 (top curve) to 18 (bottom curve), the dips induced by a spin dimer emerged and became more and more pronounced. The symbols are measured data and the curves are fitting by $\{y_0 + A_0 \exp[-(t/T_2)^4]\} \times \{y_1 + A_1 \exp[-2((t - t_c)/\omega)^2]/\omega /\sqrt {\pi /2}\}$, which consists of a smooth profile and a Gaussian dip. (b) The position of the dips is proportional to the number of pulses in the CPMG control.
 }\label{dips}
\end{figure}

The setup and experimental scheme are schematically shown in Fig.~\ref{setup}. Carr-Purcell-Meiboom-Gill (CPMG) dynamical decoupling control sequences are employed to increase the sensitivity of the NV sensor.
Figure~\ref{dips} shows the NV center spin coherence under CPMG-$N$ control. For pulse number $N>4$, we observed coherence dips on top of smooth decoherence.
The dip position $t_{\text{dip}}$ is proportional to the pulse number $N$ (Fig.~\ref{dips}b).
In the intuitive picture (Fig.~\ref{setup}c), the transition in the $^{13}$C dimer imposed an AC noise with a characteristic frequency. Therefore when the CPMG timing matches the noise characteristic frequency, the NV center spin decoherence was enhanced and therefore presented a dip\cite{Liu2011NatNano, Wrachtrup2012NatNano, Hanson2011PRL}.
As shown in previous works\cite{Liu2011NatNano, Wrachtrup2012NatNano}, the dip position is related to the precession frequency by
\begin{equation}
t_{\text{dip}}=\frac{N\pi}{\omega_{\text{ps}}^{\text{eff}}},
\end{equation}
from which we extracted the pseudo-spin precession frequency $\omega_{\text{ps}}^{\text{eff}}\approx2\pi\times (7.3 \pm 0.1)$~kHz.

\begin{figure}
\centering
\includegraphics[width=1.0\columnwidth]{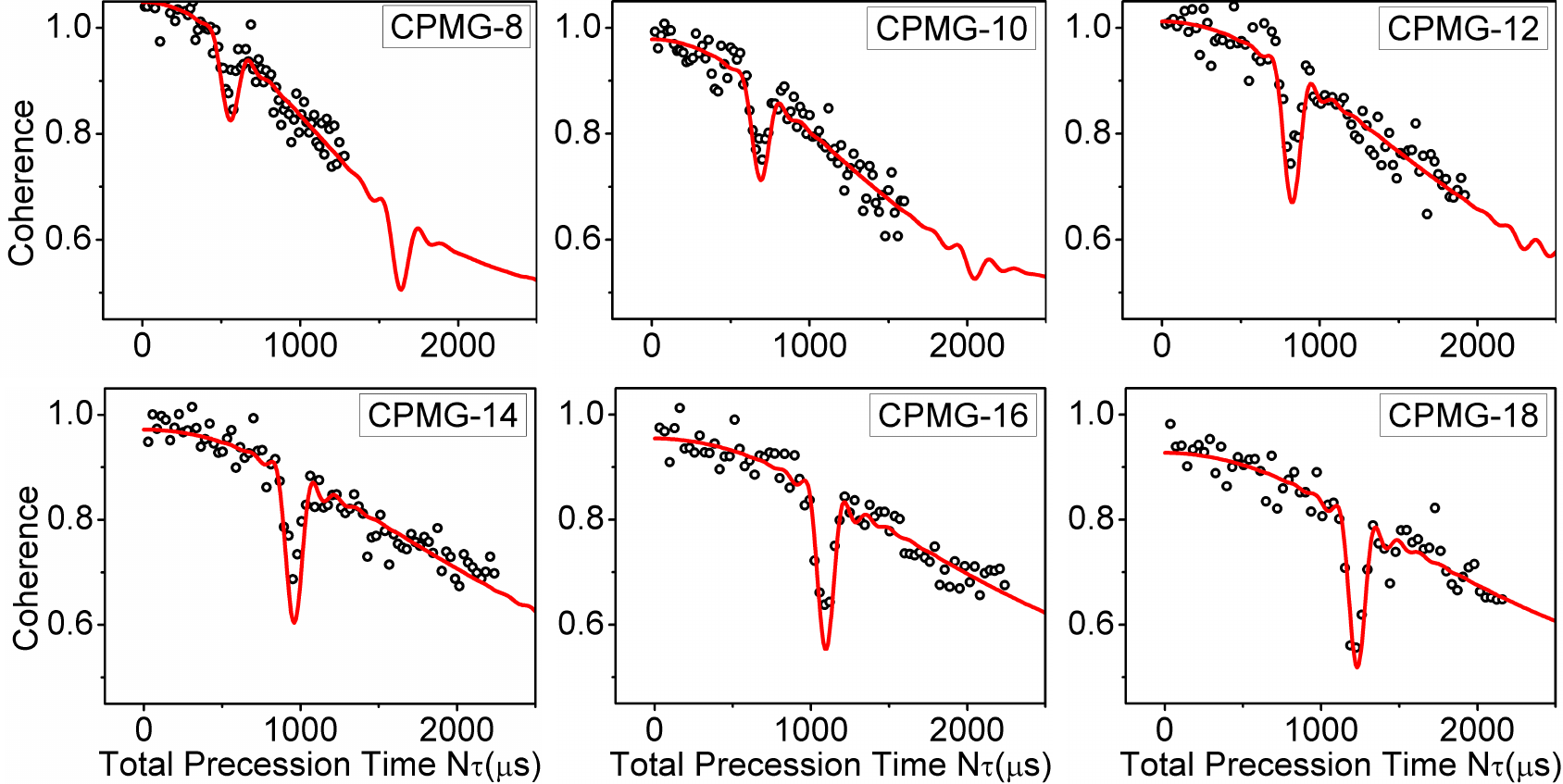}
\caption{\textbf{Characterizing interaction of a nuclear spin dimer by fingerprint-matching between measurement and simulation.}
 The figures show NV spin coherence as a function of total evolution time of CPMG dynamical decoupling control with various number of pulses ($N$=8, 10, 12, 14, 16, and 18). The experimental data (symbols) are well reproduced by simulation (lines) based on a dimer with interaction [$X$, $Z^{(m_S)}$] = [0.685, 14.6] kHz. Other nuclear spin pairs do not produce dip features matching the experimental data. The magnetic field B$_0$=1770 G in the measurement.}
    \label{simulation}
\end{figure}

The pseudo-field components $X$ and $Z^{(m_S)}$ were determined by comparing the observed dip widths and depths under various control pulse number $N$ with the calculated electron spin coherence using the pseudo-spin model.
Because the nuclear spin positions are discrete on diamond lattice,
the dipolar interaction takes discrete values $X=2.1~\text{kHz}$ for a dimer in $[111]$ direction,
$X=0.685~\text{kHz}$ for a dimer in other C-C bond directions, and $X<0.3$~kHz for nuclear spins with larger distance.
The calculation shows that only parameters $[X, Z^{(1)}] = [0.685, 14.6]$~kHz
yield full agreement with the experiment data shown in Fig.~\ref{simulation}.
With this we conclude that the coherence dip was caused by a dimer along non-[111] direction (SI, Sec.~3.1).

We characterized the coupling between the dimer and the NV center electron spin by measurement for various magnetic fields.
The $Z^{(m_S)}$ component of the pseudo-field consists of two part $Z^{(m_S)} = Z^{(m_S)}_{\parallel} + Z^{(m_S)}_{\perp}(B_0)$.
The parallel part $Z^{(m_S)}_{\parallel}=A_{1,\parallel}-A_{2,\parallel}$ is the difference of the hyperfine fields projected to the applied magnetic field direction.
The transverse part $Z_{\perp}^{(m_S)}(B_0) = \frac{(A_{1,\perp}^{(m_S)})^2-(A_{2,\perp}^{(m_S)})^2}{\gamma _CB_0}$ comes from the perturbative frequency shifts of the two spins induced by virtual flipping via the polarized states $\vert \uparrow\uparrow\rangle$ and $\vert \downarrow\downarrow\rangle$\cite{Liu2012PRB}.
Notice that the parallel part is opposite for $\vert m_S=1\rangle$ and $\vert m_S=-1\rangle$ states, but the transverse part is the same.
The presence of the transverse part $Z_{\perp}^{(m_S)}(B_0)$ causes different dip positions for the transitions
$\vert m_S=0\rangle\leftrightarrow\vert +1\rangle$ and $\vert m_S=0\rangle\leftrightarrow\vert -1\rangle$. The dip position difference
$\delta t_\text{dip}$ depends on the magnetic field.
In the case of $Z^{(m_S)}_{\parallel} \gg Z^{(m_S)}_{\perp}(B_0)$, which is the case in our experiment,
\begin{equation}
\delta t_{\text{dip}} \approx \frac{2N {\color{red}}}{(Z_{\parallel}^{(1)})^2}\frac{(A_{1,\perp})^2-(A_{2,\perp})^2}{\gamma_{\text{C}}B_0}.
\end{equation}
In the strong magnetic field limit , $\delta t_{\text{dip}} \rightarrow 0$,
and $Z_{\parallel}^{(1)}=14.6$~kHz is extracted from $\omega_{\text{ps}}^{\text{eff}}$, as shown in Fig.\ref{simulation} for $B=1770$~Gauss.
The magnetic field dependence of $\delta t_{\text{dip}}$ gave the difference of the transverse hyperfine coupling $\delta A_{\perp}^2=(A_{1,\perp})^2-(A_{2,\perp})^2 = (21.7~\text{kHz})^2$ (SI, Sec.~3.2).
With the values of $Z^{(1)}$ and $\delta A_{\perp}^2$, we determined the dimer was located about 1~nm away from the NV center (SI, Sec.~4). The field dependence of the dip features also excludes electron spin clusters as the the cause of the coherent modulation (SI, Sec.~5).

In summary, we have observed and characterized nuclear spin interaction as weak as 685~Hz within a single two-nucear-spin cluster in diamond.
By decoherence measurement of the NV center spin under multi-pulse dynamical decoupling control and various magnetic fields and by comparing the data with calculations, we determined the orientation, distance between the two nuclear spins, and distance from the NV center of the dimer, with atomic-scale resolution.
These results demonstrate that the dynamical decoupling technique is promising for deriving the
interaction between nuclear spins and therefore for structure analysis of single molecules on diamond surface using shallow NV centers.

\begin{methods}
The NV sensor in $<100>$ face bulk diamond was mounted on a typical ODMR confocal setup which was synchronized together with the microwave bridge by multi-channels pulse generator (Spincore, PBESR-PRO-500).  The nitrogen concentration was less than 5ppb and the abundance of $^{13}$C was naturally equal to 1.1\%. The pump 532nm green laser and phonon sideband fluorescence (wavelength, 650-800 nm) went through the same oil objective (Olympus, PLAPON 60XO, NA 1.42). Before the pump laser went into the objective, it was forced to go double through one AOM (ISOMET, power leakage ratio $\sim$1/1000), to preserve $T_1$ of NV sensor from the leakage laser effect. Fluorescence intensity, collected by avalanche photodiodes (Perkin Elmer, SPCM-AQRH-14) with a counter card, was used to monitor the magnetic signal on the NV sensor. The adjustable external magnetic field (0$\sim$2000 Gauss) created by the electrical coil (GMW model 5201, 3ppm/hour stability) was aligned by monitoring the diversification of fluorescence. Pulsed microwave was carried by a 20$\mu$m copper wire. The microwave phase was continuously adjusted by mechanical phase shifter and fixed to X and Y phase by network analyser (Agilent). High order CPMG-$N$ sequences were used to amplify the weakly signal $Z^{(m_S)}$ from remote nuclear spin pair, $Z^{(m_S)} \sim \text{kHz} \ll \text{MHz} \sim 1/T_2^*$, in which, $T_2^*$ was roughly 1 $\mu$s in our sample. CPMG-12 on the different transitions 0 $\leftrightarrow \pm1$ under various magnetic field were employed to characterize the coupling vector and distinguish the high field threshold. Based on the experiment condition, the pseudo spin model was used to analyze the experiment result.

%
%

\end{methods}

\begin{addendum}
\item [Acknowledgements]
This work was supported by the National Key Basic Research Program of China (Grant No. 2013CB921800), the National Natural Science Foundation of China (Grant Nos. 11227901, 91021005, 10834005, 11028510), the 'Strategic Priority Research Program (B)' of the CAS (Grant No. XDB01030400), the Fundamental Research Funds for the Central Universities, Hong Kong Research Grants Council (N\_CUHK403/11, 402410 \& HKU8/CRF/11G),  the Chinese University of Hong Kong Focused Investments Scheme and 1000 Plan Program for Young Talents.

\item[Author contributions]
J.D. and R.B.L. proposed the idea. J.D. and F.S. designed the experimental proposal. P.W., X.K. and F.S. prepared the experimental set-up. F.S., F.K. and X.K. performed the experiments. J.D. supervised the setup and experiments. N.Z. and F.S. carried out the calculation. F.S., N.Z., R.B.L. and J.D. wrote the paper.
All authors analysed the data, discussed the results and commented on the manuscript.

\item[Competing Interests] The authors declare that they have no competing financial interests.
\item[Additional information]
 Supplementary information accompanies the paper on http://www.nature.com/nnano.
Correspondence and requests for materials should be
addressed to J.D.
\end{addendum}

\begin{figure}
\centering
\includegraphics[width=0.75\columnwidth]{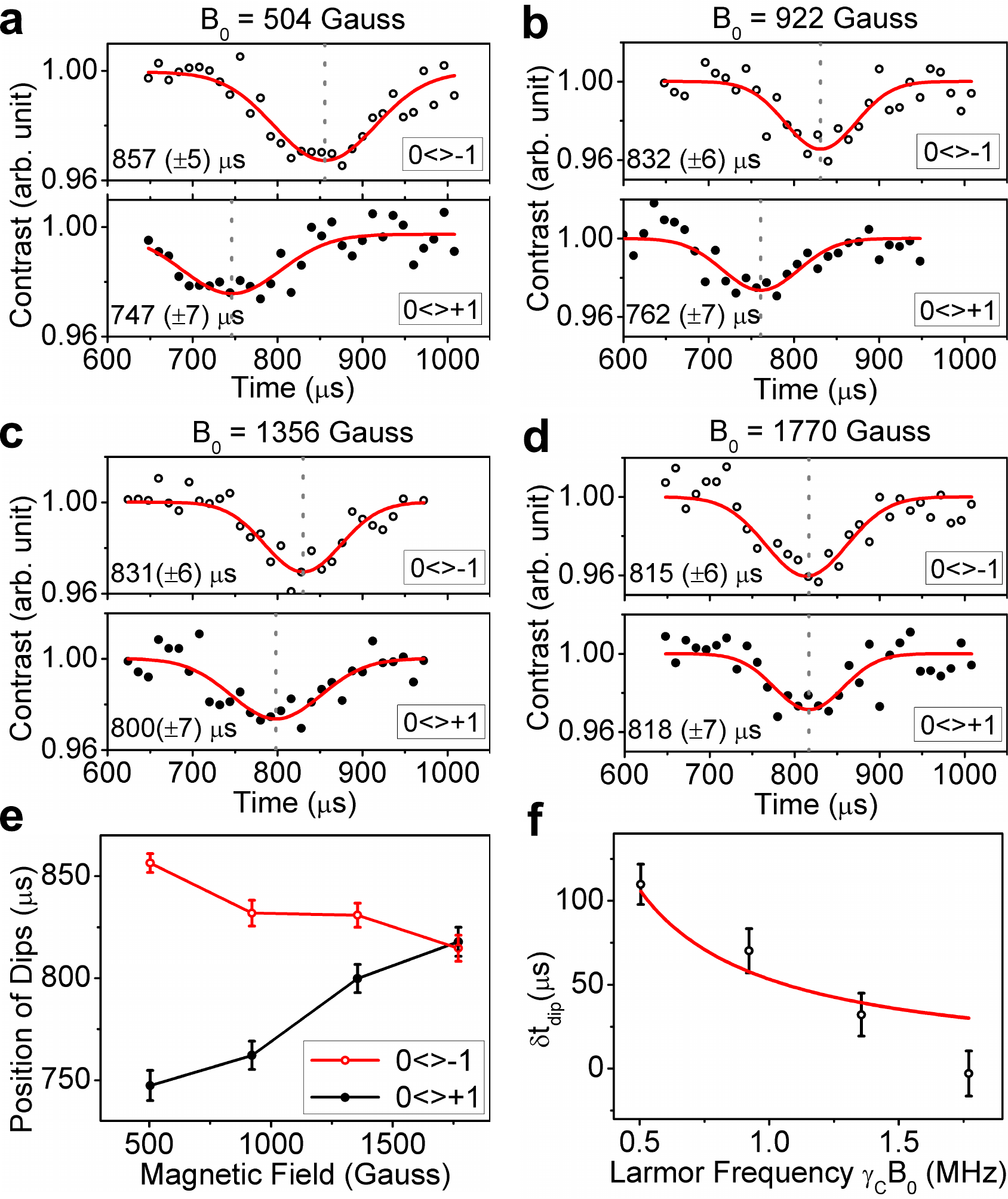}
    \caption{\textbf{Magnetic field dependence of the dip features for different transitions in the NV center spin decoherence.}
   (a-d)The upper and lower panels in each part show the CPMG-12 decoherence of the transitions $|0\rangle\leftrightarrow|\text{+}1\rangle ~\text{and}~ |0\rangle\leftrightarrow |\text{-}1\rangle$, respectively. The magnetic field for (a)-(d) was in turn 504, 922, 1356, and 1770 Gauss. (e) Field dependence of the dip positions for different NV center spin transitions. (f) Comparison between the measured (symbols) and calculated (lines) positions of the dips as functions of the magnetic field.
 }
    \label{field}
\end{figure}

\end{document}